\documentclass[]{raa}            
\usepackage{graphicx,times}      
\usepackage{natbib}
\usepackage{url}
\usepackage{longtable}

\def\kms{$\rm{km~s}^{-1}$}

\begin{document}

\title{Automatic Determination of Stellar Atmospheric Parameters and Construction of Stellar Spectral Templates
   of the Guoshoujing Telescope (LAMOST)}

   \volnopage{ Vol.0 (2011) No.0, 000--000}
   \setcounter{page}{1}

   \author{Yue Wu
      \inst{1,2,3,4}
   \and A-Li Luo
      \inst{1,3,5}
   \and Haining Li
      \inst{1,3}
   \and Jianrong Shi
      \inst{1,3}
   \and Philippe Prugniel
      \inst{2}
   \and Yanchun Liang
      \inst{1,3}
   \and Yongheng Zhao
      \inst{1,3,5}
   \and Jiannan Zhang
      \inst{1,3}
   \and Zhongrui Bai
      \inst{1,3}
   \and Peng Wei
      \inst{5,1,3}
   \and Weixiang Dong
      \inst{5,1,3}
   \and Haotong Zhang
      \inst{1,3}
   \and Jianjun Chen
      \inst{1,3}
   }

   \institute{National Astronomical Observatories, Chinese Academy of Sciences,
             20A Datun Road, Chaoyang District, Beijing, 100012, China; \\
             {\it wuyue@lamost.org, lal@lamost.org} \\
        \and
        Universit\'e de Lyon, Universit\'e Lyon 1, Villeurbanne, F-69622, France;
        CRAL, Observatoire de Lyon, CNRS UMR~5574, 69561 Saint-Genis Laval, France; \\
        \and
        Key Laboratory of Optical Astronomy, NAOC, Chinese Academy of Sciences; \\
        \and
        Graduate University of the Chinese Academy of Sciences, 19A Yuquan Road,
        Shijingshan District, Beijing, 100049, China \\
        \and
        Shandong University at Weihai, 180 Wenhuaxi Road, Weihai, Shandong Province, 264209, China \\
   }

   \date{Received [2010] [month] [day]; accepted [2010] [month] [day]}

\abstract{ A number of spectroscopic surveys have been carried out 
or are planned to study the origin 
of the Milky Way. 
Their exploitation requires reliable automated
methods and softwares to measure the fundamental parameters
of the stars.
Adopting the ULySS package, we have
tested the effect of different resolutions and signal-to-noise
ratios (SNR) on the measurement of the stellar atmospheric parameters (effective
temperature $T_{\rm{eff}}$, surface gravity log~$g$, and metallicity
[Fe/H]). We show that ULySS is reliable to determine
these parameters with medium-resolution spectra
(R$\sim$2000). Then, we applied the method to
measure the parameters of 771 stars selected in
the commissioning database of the
Guoshoujing Telescope (GSJT). The results were compared with the
SDSS/SEGUE Stellar Parameter Pipeline (SSPP), and we derived precisions
of 167\,K, 0.34\,dex, and 0.16\,dex for
$T_{\rm{eff}}$, log~$g$ and [Fe/H] respectively. Furthermore, 120 of
these stars are selected to construct the primary
stellar spectra template library (Version 1.0) of GSJT, and will
be deployed as basic ingredients for the GSJT automated parametrization
pipeline.
 }

   \authorrunning{Y. Wu et al. 2011} 
   \titlerunning{Determining Atmospheric Parameters and Constructing Stellar Spectral Templates for GSJT}
   \maketitle

\section{Introduction}
 \label{sect:intro}

The origin and evolution of the Milky Way are key subjects 
in modern astrophysics. To explore these subjects, it is 
essential to understand the intrinsic properties, such as 
stellar masses, ages, chemical abundances, and kinematics 
for statistically significant samples of stars in the 
Galaxy. They will be used to match the structure and 
evolution of the Milky Way to the current generation of 
galaxy formation models 
\citep{wyse06,juric08,ive08,bond10,jofre11}.

Thanks to the development of astronomical technology
and instruments, e.g., multi-fiber and multi-aperture spectrographs,
adaptive optics, etc., large survey projects have become possible. These
include the successful projects such as the Sloan Digital Sky
Survey (SDSS, York et al. 2000), the follow-up Sloan Extension
for Galactic Understanding and Exploration (SEGUE, Yanny et al.
2009) and SEGUE-2 \citep{rock09}, as well as ongoing surveys such as the Radial
Velocity Experiment (RAVE, Steinmetz et al. 2006, Zwitter et al.
2008, Siebert et al. 2011), the Guoshoujing Telescope
(GSJT\footnote{\url{http://www.lamost.org/website/en/}}, formerly 
named as the Large sky Area Multi-Object fiber Spectroscopic
Telescope\,-\,LAMOST) survey,
SDSS-III\footnote{\url{http://www.sdss3.org/}},
and a number of surveys which are being planned, e.g., Galactic (GAIA, Perryman et
al. 2001), APOGEE \citep{schiavon10} and HERMES \citep{wylie10}.
Unprecedentedly large spectroscopic databases of Galactic stars
are becoming available.

This data avalanche calls for new analysis tools 
for automated and efficient parameterizations of stellar spectra.
The basic atmospheric parameters, $T_{\rm{eff}}$, log~$g$, [Fe/H], are
some of the characteristics that the analysis pipelines of the spectroscopic
surveys shall determine. 
Many research teams
have made great efforts in this field, aiming to extract efficiently and reliably the maximal
astrophysical information, especially
to determine the atmospheric parameters over 
wide ranges of $T_{\rm{eff}}$, log~$g$, and [Fe/H]. Numerous methods
have been developed in order to extract those atmospheric parameters
from medium-resolution stellar spectra in a fast, automatic, objective
fashion. Generally, these approaches can be sorted into two main
categories, the minimum distance method (MDM), and the non-linear
regression method, commonly called Artificial Neural Network or ANN.
Except for the above two main categories, there are other methods,
like correlations between broadband colors or the strength of
prominent metallic lines and the atmospheric parameters 
(Beers et al. 1999; Wilhelm et al. 1999; Cenarro et al. 2001ab; 
Alonso et al. 1996ab; Alonso et al. 1999ab; Ivezi\'c et al. 2008; 
An et al. 2009; Arnad\'ottir et al. 2010).

MDM has been widely used in the past, not only for the automatic
parametrization of $T_{\rm{eff}}$, log~$g$, [Fe/H], radial velocity
etc., but also for the spectral classification. The solution is based
on the best matching template spectrum according to the shortest
distance in an N-dimensional data space, where N is the number of
selected spectral features. To use this kind of method, first,
we need to construct a stellar spectral template library with which
accurate parameters have been previously determined via traditional methods.
The $\chi^2$ minimization, the cross-correlation, the weighted
average algorithm, and the k-nearest neighbor (KNN) method, etc., are
various specific cases of the MDM. There are many representative works
assigned in this category. For example, Katz et al. (1998) developed
the TGMET software to derive parameters by a direct comparison with
a reference library of stellar spectra. This work was later updated
by Soubiran et al. (2000, 2003) to an internal accuracy of 86\,K,
0.28\,dex and 0.16\,dex respectively for $T_{\rm{eff}}$, log~$g$ and
[Fe/H] for F, G and K stars with SNR of 100, and accuracy of 102\,K,
0.29\,dex and 0.17\,dex at SNR\,=\,10. In Zwitter et al. (2008), the
parameters are derived by a penalized $\chi^2$ method using an
extensive grid of synthetic spectra calculated from the latest
version of Kurucz stellar atmospheric models. There are many other
typical works in this category, including Zwitter et al. (2005),
MATISSE \citep{MATISSE}, Allende Prieto et al. (2006), Shkedy et al.
(2007), Jofr\'e et al. (2010), as well as the SSPP (Lee et al. 2008a),
in which they employed various techniques in integrated approaches
containing both the MDM and ANN methods.

Meanwhile, a number of early studies have demonstrated that ANN 
methods can be robust and precise for parametrization, by
providing a functional mapping between the spectra as its inputs and
the parameters as its outputs. The optimal mapping is found by
training the network (i.e. setting its weights) on a set of either
pre-classified observed stellar spectra or synthetic stellar
spectra (both used as templates). Re Fiorentin et al. (2007) succeeded
to derive atmospheric parameters from medium-resolution 
stellar spectra using non-linear regression models which can be
trained alternatively on either real observations or synthetic
spectra. By comparing with the SDSS/SEGUE data, they reached RMS
deviations on the order of 150\,K in $T_{\rm{eff}}$, 0.35\,dex in
log~$g$, and 0.22\,dex in [Fe/H]. Similar efforts can be found in
SSPP (Lee et al. 2008a),
Bailer-Jones et al. (1997),
Bailer-Jones (2000),
Willemsen et al. (2005),
Zhao et al. (2006), and
Manteiga et al. (2010).

For both techniques, a comprehensive set of reference spectra
with known atmospheric parameters is crucial.
Such {\it stellar spectral libraries} with medium to high
resolution and good coverage of the Hertzsprung-Russell diagram
are also an essential tool in several areas of astronomy, including:

\begin{enumerate}
  \item Automated classification and parametrization of large
volumes of stellar spectral data, especially those collected by ongoing
large spectroscopic surveys;
  \item Spectral synthesis of stellar populations of galaxies,
e.g. Le Borgne et al. (2004), Percival \& Salaris (2009), Prugneil
et al. (2007a), Koleva et al. (2007), Vazdekis et al. (2010), and Chen
et al. (2010);
  \item Derivation of radial velocities via cross-correlation
against the best matched templates, e.g. Tonry \& Davis (1979),
Valentini \& Munari (2010);
  \item Calibration of spectroscopic line/band classification
criteria;
  \item Calibration of photometric indices.
\end{enumerate}

The larger quantity and better quality of newly obtained spectra have 
promoted improvements of these libraries, and nowadays the empirical and
synthetic stellar spectral libraries are developed in parallel.
For the former ones, the spectra are the real observations ideally
collected with a single instrumentation and setup. Some popular
empirical libraries are:
ELODIE \citep{PS01, PS04, prug07b};
UVES-POP \citep{uves03};
STELIB \citep{stelib03};
CFLIB (The INDO-US library of Coud\'e-feed Stellar Spectra) \citep{val04};
MILES \citep{milesi06, milesii07};
NGSL \citep{heap07}.
For the latter ones, the spectra are calculated with
a stellar model atmosphere, e.g. ATLAS9 \citep{kurucz93},
Kurucz 2003 \citep{castelli03}, MARCS \citep{marcs08}, Ian Short \&
Hauschildt (2010), and lists of atomic and molecular lines, e.g.
Cayrel et al. (2001), Barbuy et al. (2003). Some important 
synthetic libraries are:
BaSeL \citep{BaSeL02};
a grid of synthetic spectra and indices for Fe\,I\,5270, Fe\,I\,5335, Mg\,Ib and
$\rm{Mg_{2}}$ as a function of stellar parameters and [$\alpha$/Fe]
\citep{barbuy03};
a library of high-resolution Kurucz spectra in the range $\lambda$\,
3000-10,000~\AA{} \citep{murphy04};
an extensive library of 2500\,-\,10,500~\AA{} synthetic spectra
\citep{munari05};
a high-resolution stellar library for evolutionary population
synthesis \citep{martins05};
a library of high resolution synthetic stellar spectra from
300\,nm to 1.8\,$\mu$m with solar and $\alpha$-enhanced composition
\citep{coelho05};
UVBLUE \citep{rodri05};
synthetic stellar and SSP libraries as templates for Gaia simulations
\citep{sordo10}.
All these libraries widely differ in the number of stars, the
calibration quality, wavelength intervals, spectral resolution,
range of atmospheric parameters, etc.

The choice between the use of empirical and synthetic libraries is a
subject of debates in the literature. The advantage of the observed
libraries is that they are {\it real}, so avoiding any simplifying
assumptions in the synthetic spectra calculation \citep{munari01},
but the limitation is the inability to extrapolate to abundance
patterns differing from those of the input stars. The
synthetic libraries can overcome this limitation by
providing a more uniform coverage in the parameter space, but the
problem remains that they depend on model atmospheres, with potential
systematic uncertainties. For example, Kurucz models
\citep{kurucz93, castelli03} assume local thermodynamic
equilibrium, which is known to break down in a number of regimes
(e.g. for very hot stars). Besides, $T_{\rm{eff}}$, log~$g$, and
[Fe/H] do not uniquely describe a true spectrum, hence advanced
models that are sensitive to different abundance ratios and
concerning, e.g., chromospheres/dust are necessary.

The GSJT \citep{su98, xing98, zhao00, zhu06} is
a unique astronomical instrument for large area spectroscopic
survey. It can simultaneously obtain spectra of 4000 celestial
objects due to its special multi-fiber design. Since 2009,
GSJT has initiated its commissioning stage, which mainly consists 
of tests on its  reliability and stability,
and adjustments for optimized performance and operation. 
One of the major scientific aims of the GSJT project
is to study the
formation and evolution of our Galaxy.
The observed stars are expected to include almost all 
evolutionary stages and a wide range of masses in the 
Hertzsprung-Russell diagram.

To achieve the above scientific goal, a prerequisite is the
quality of the GSJT 1D stellar parameter pipeline (Luo et al.
2008, using a non-linear regression method), which is responsible for
automatic spectral classification and parametrization.
Obviously, a major element is the database of stellar spectra templates. 
We plan to take a representative set of
observed stellar spectra across the entire parameter space, determine
accurate atmospheric parameters, and adopt
these spectra to train the parameterizers. 
We will build this database with GSJT spectra in order to supress
the effects due to observational signatures.
We will assemble a first version of this template library using the
data acquired during the commissioning period.
In this paper, we employ ULySS to automatically
determine the stellar atmospheric parameters, test the accuracy
of the measurements, and present a first version of the  GSJT template
library. The observation and data reduction are
described in Sect. 2. Methods for the determination of the stellar
atmospheric parameters and their validation are presented in Sect. 3.
The adopted parameters of the selected GSJT commissioning spectra
are described in Sect. 4. Our proposed stellar spectral templates
are provided in Sect. 5, and our summary and futher prospects are
discussed in Sect. 6.

\section{Observation and Data Reduction}
 \label{sect:obs&red}

Since at the time our sample was assembled, 
the GSJT was still in its engineering commissioning phase
and subject to instrumental instability, 
in order to maintain homogeneity as much as possible, we
adopted the spectra observed in one GSJT commissioning field on
February 13, 2010. Most of the observations are field stars around
M67. The input targets were taken from the UCAC3 \citep{zachar10} 
catalog with $R$ band magnitudes brighter than 16.5. The 
observations were done with double exposures of 30 minutes and 20 
minutes respectively, and a seeing of $\sim$\,3.3'' (FWHM of the 
CCD image of the guide star)\footnote{At that time, the dome 
seeing conditions of the GSJT were not ideal. At present, work is 
ongoing to improve this situation by adjusting the ventilation 
and cryogenic systems of the interior of the dome and the 
telescope tube. In the near future, we will continue to correct 
the comparatively large temperature difference of the Ma mirror, 
Mb mirror and the focal plane of the GSJT, e.g. by employing 
refrigeration on these locations.}.

\begin{figure}
\centering
\includegraphics{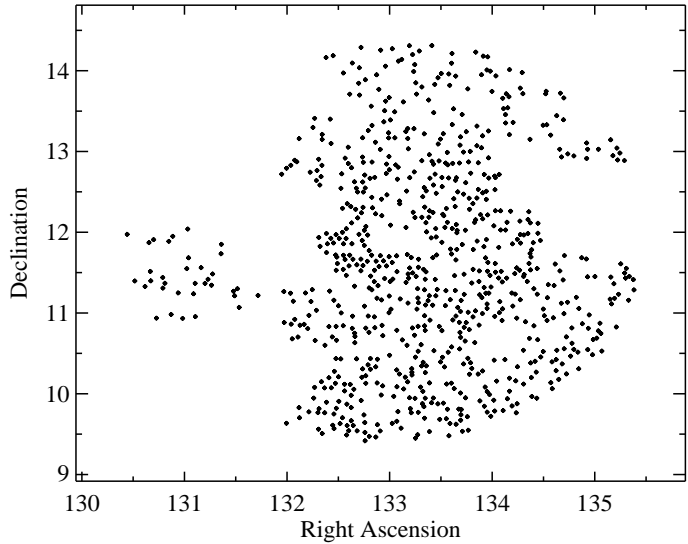}
\caption {Projection of the selected 771 GSJT stellar observations
in equatorial coordinates.}
 \label{fig.radec}
\end{figure}

The resolving mode R\,=\,2000 was adopted. It was achieved by setting 
the slit width to half of the diameter of the fibers, which inevitably
leads to loosing 39\% of the incoming light (the percentage of the cross
section area which was blocked). This sacrifice is the price to enhance 
the resolution that would be around R\,=\,1000 otherwise.
The spectral coverage was between 370\,-\,590\,nm and 570\,-\,900
\,nm for the blue and red arms of the spectrographs, respectively. Four 
thousand fibers were allocated to 16 spectrographs from No.1 through 
No.16. However, due to the design of the GSJT, fibers closer to the
center of the focal plane suffered less vignetting and hence showed
better image quality. Therefore, our selection was biased against targets 
located at the outer region of the focal plane. In addition, the 
present relatively low optical efficiency, inaccuracy in fiber positioning, 
and uncertainties in the flat-fielding and sky-subtraction lead to a 
large number of low-quality spectra. We further excluded spectra with 
SNR less than 10. Out of the 4000 observed objects, the above selection 
yielded a final sample of 771 stars; these data will be analyzed and 
adopted as basic ingredients to construct the preliminary stellar 
spectral template library for GSJT. Note that data of some selected 
objects showed only one exposure with good quality while others were 
acceptable in both exposures. For the latter case, the two observations 
were combined to enhance the SNR. The coordinates and $R$ band 
magnitudes of the selected objects are listed in Table~\ref{tb.meas} 
(Only 5 objects were shown. The full table is only available in the
electronic version). $R$ band magnitudes of the selected sample are
between 7.29 and 16.37, and the spatial distribution is shown in
Fig.~\ref{fig.radec}.

The raw data were reduced with the GSJT standard 2D pipeline
\citep{luo04}, including bias subtraction, cosmic-ray removal,
1D spectral trace and extraction, flat-fielding, wavelength
calibration, and sky subtraction. Since the flux calibration 
was not yet reliable, this step was not implemented in our 
sample spectra.

\section{Application and Validation of ULySS to Atmospheric Parameter Determination with stellar spectra at R\,=\,2000}
 \label{sect:method}

In this work we employed ULySS\footnote{available
at: \url{http://ulyss.univ-lyon1.fr}} (Universit\'e de Lyon
Spectroscopic analysis Software, Koleva et al. 2009a) to analyze the
selected spectra. This package
is written in the Interactive Data Language (IDL) and enables full
spectral fitting for various astrophysical applications, including the
determination of (i) stellar atmospheric parameters, e.g., the TGM 
case, T, G and M respectively represent the effective temperature, 
the surface gravity and the metallicity \citep{wu11a,prug11a,koleva09a}, 
and (ii) the star formation and metal enrichment history of galaxies, 
e.g., SSP (simple stellar population) case \citep{michi07,koleva09b,
koleva09c,bouch10,maka10,smirnova10,sharina10}. It minimizes 
$\chi^2$ between an observed spectrum and a model spectrum, and the 
fit is performed in the pixel space (evenly sampled in logarithm of 
wavelength). 
The method
determines all the free parameters in a single fit in order to
properly handle the degeneracy between the temperature and the
metallicity. This method has already been successfully tested
in Wu et al. (2011a). In this section, we will briefly
introduce the usage of ULySS to determine the stellar
atmospheric parameters using
medium-resolution spectra (e.g. obtained by SDSS and GSJT).
We evaluate the effect of the resolution and SNRs.

\subsection{ULySS Method}

In ULySS, an observed spectrum is fitted against a model expressed
as a linear combination of non-linear components, optionally
convolved with a line-of-sight velocity distribution (LOSVD) and
multiplied with a polynomial function. A component is a non-linear
function of some physical parameters, e.g., $T_{\rm{eff}}$, log~$g$
and [Fe/H] for the TGM case. The multiplicative polynomial is meant to
absorb errors in the flux calibration, Galactic extinction or any
other source affecting the shape of the spectrum. It replaces the
prior {\it rectification} or normalization to the pseudo-continuum
that other methods require, as in Valentini \& Munari (2010). This
model is compared to the data through a non-linear least-square
minimization. The minimization issue can be written as:

\vspace {0.5cm} $\rm Obs(\lambda) = P_{n}(\lambda) \times
[\,\rm{TGM}(\,T_{\rm{eff}},log~$g$,[Fe/H],\lambda) \otimes
G(\,v_{sys},\sigma)]$, \vspace {0.5cm}

where $\rm Obs(\lambda)$ is the observed stellar spectrum
($\lambda$ is the logarithm of the wavelength), $\rm
P_{n}(\lambda)$ a development in Legendre polynomials of order n, and $\rm
G(v_{sys},\sigma)$ is a Gaussian broadening function characterized
by the systemic velocity $v_{sys}$, and the dispersion $\sigma$. The
free parameters of the minimization process are the three parameters
of the TGM function ($T_{\rm{eff}}$, log~$g$ and [Fe/H]), the two
parameters of the Gaussian ($v_{sys}$ and $\sigma$), and the
coefficients of $P_{n}$. $v_{sys}$ absorbs the imprecision of the
cataloged radial velocity of the stars which were used to reduce
them in the rest frame; $\sigma$ encompasses both the instrumental
broadening and the effects of stellar rotation. 

\begin{figure}
\centering
\includegraphics[width=13cm]{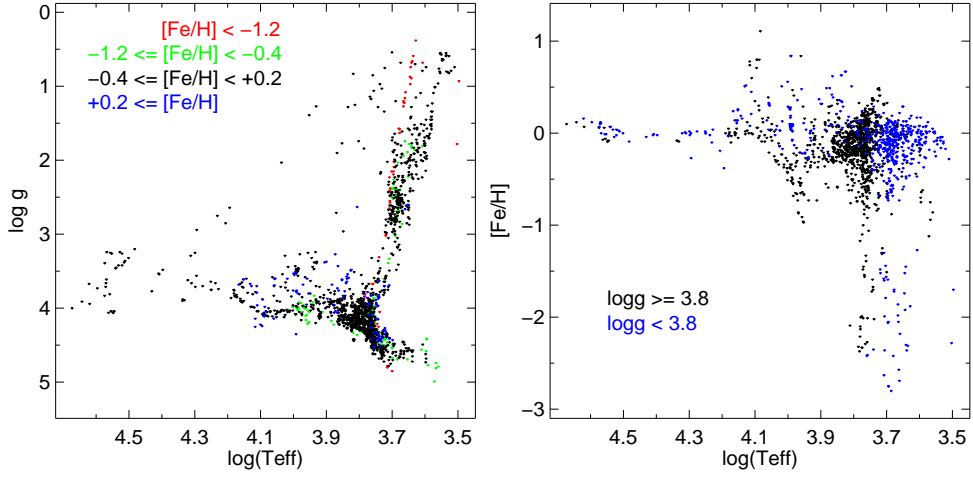}
\caption {Distribution of the ELODIE (version 3.2) interpolator
inverted atmospheric parameters in the log($T_{\rm{eff}}$) - log~$g$
plane (left), and in the log($T_{\rm{eff}}$) - [Fe/H] plane (right).
In the left plot, the color of the symbols distinguishes different
metallicity classes. On the right, the dwarfs are shown in black
and the giants in blue.}
\label{fig.dist_eloI}
\end{figure}

The TGM function is an {\it interpolator} of the ELODIE
library \citep{PS01, PS04, prug07a} version 3.2 \citep{wu11a, prug11b},
which has a wavelength coverage of 3900\,-\,6800~\AA, and a
resolution version of R\,=\,10,000.
The interpolator consists of polynomial expansions of each
wavelength element in powers of log($T_{\rm{eff}}$), log~$g$, [Fe/H]
and $f(\sigma)$ (a function of the rotational broadening
parameterized by $\sigma$, the standard deviation of a Gaussian).
Three sets of polynomials are defined for three temperature ranges
(roughly matching OBA, FGK, and M types) with important overlap between
each other where they are linearly interpolated. For the FGK and M
polynomials, 26 terms are adopted; for OBA, 19 terms are used.
The coefficients of these polynomials were fitted over the $\sim$2000
spectra of the library, and the choice of the terms of the
$T_{\rm{eff}}$ limits and of weights were fine tuned to minimize the
residuals between the observations and the interpolated spectra. The
interpolator based on the latest ELODIE library (version 3.2)
provides valid inverted atmospheric parameters covering
3100$\sim$59000\,K in $T_{\rm{eff}}$, 0.00$\sim$5.00\,dex in log~$g$,
and -2.80$\sim$1.00\,dex in [Fe/H], with the parameter space
coverages shown in Fig.~\ref{fig.dist_eloI} \citep{prug11b}.

\subsection{Validation}

In the ULySS TGM case, the model generated based on the ELODIE
library\footnote{\url{http://www.obs.u-bordeaux1.fr
/m2a/soubiran/elodie_library.html}} is with FWHM\,=\,0.58~\AA~, R\,$\sim$\,10000.
When ULySS performs the fitting, the model is adjusted at the same
resolution and sampling as the input observation. Since ULySS has
never been used to fit spectra at a resolution as
low as that of GSJT spectra, i.e., R\,=\,2000, we tested the effect
of lower resolutions as well as different SNRs by using the CFLIB
library \footnote{\url{http://www.noao.edu/cflib/}} and some
SDSS/SEGUE\footnote{\url{http://www.sdss.org/segue/}} spectra.

\subsubsection{Validation with the Coud\'e-feed Stellar Spectral Library}

The CFLIB library \citep{val04} contains 1273 stars obtained using
the 0.9m coud\'e feed telescope at Kitt Peak National Observatory;
it has a wide wavelength coverage from 3460\,-\,9464~\AA, and a
resolving power of R$\sim$5000 (FWHM$\sim$1.2\AA). 
This sample of 1273 stars was
selected to cover a broad regime of the three atmospheric parameters
($T_{\rm{eff}}$, log~$g$ and [Fe/H]), as well as the spectral type.
In Wu et al. (2011a), ULySS was used to homogeneously
measure the atmospheric parameters for this library, and the determinations
were extensively compared with the results from many other previous
studies based on high resolution spectra and traditional estimation 
methods. For the 958 F, G, and K type stars of this library, the 
precision of determination for $T_{\rm{eff}}$, log~$g$ and [Fe/H] are 
43\,K, 0.13\,dex and 0.05\,dex respectively, and no significant 
systematic biases were found for the three parameters.    

\begin{figure}
\includegraphics{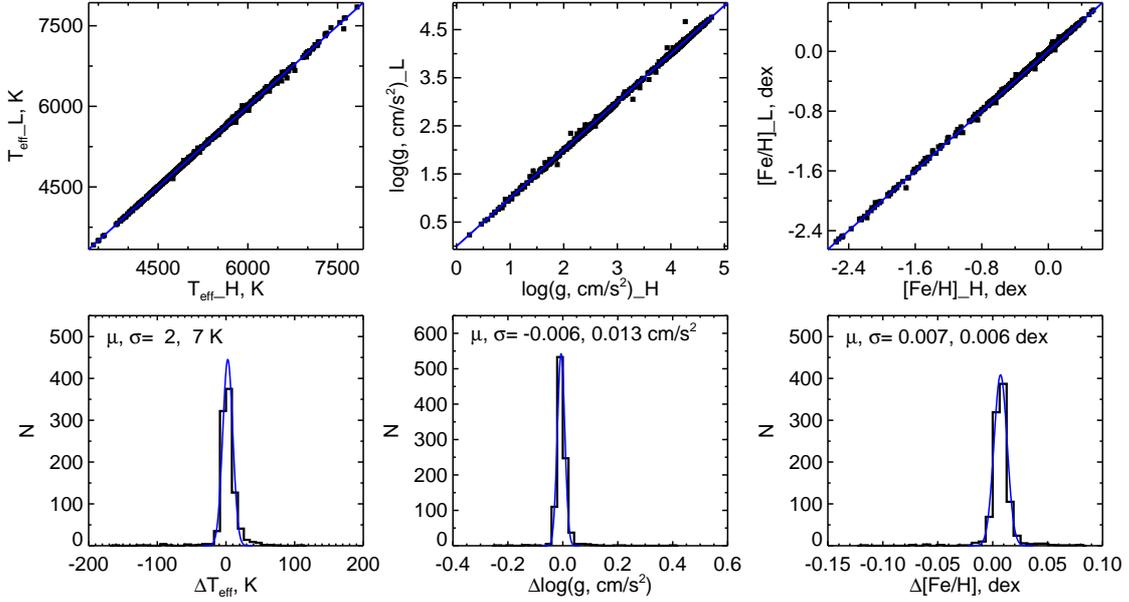}
\caption {Comparison of the CFLIB library 958 F, G and K type stars' 
ULySS determined atmospheric parameters, between fitting the original CFLIB
observations (R$\sim$5000, upper panels abscissas) and those after
decreasing the resolution to the same level of the GSJT (R$\sim$2000,
upper panels ordinates). The blue solid lines in the upper panels are
the 1 to 1 ratios. The histograms of the discrepancies between the
two series (ordinate values minus abscissa values) are shown in the
lower panels, with a fitted Gaussian in thinner blue line overplotted. 
The offset and 1\,$sigma$ standard deviation for each parameter 
between the two series are listed in the lower panels respectively.}
 \label{fig.cftest}
\end{figure}

To test the effect of the lower resolution,
we degraded the CFLIB spectra to match the resolving power of
the GSJT observations (resampling the original CFLIB spectra to 36\,\kms
per pixel on a logarithmic scale), and the determined atmospheric
parameters were then compared with those published in Wu et al. (2011a).
The results are presented in
Fig.~\ref{fig.cftest}. It is shown that for $T_{\rm{eff}}$, log~$g$,
and [Fe/H], the offsets are not significant (3\,K, -0.006\,dex
and 0.007\,dex respectively); the 1$\sigma$ scatters are also 
negligible (7\,K, 0.013\,dex and 0.006\,dex respectively). This 
demonstrates that the sacrifice in resolution of the input observation, 
as low as R$\sim$2000, will not affect the final accuracy of the 
derived atmospheric parameters, hence we can trust the results and 
apply ULySS to the GSJT observations.

\subsubsection{Validation with SDSS Spectra: High Resolution Spectra and the SNR Test}

Since the SDSS spectroscopic survey has a similar
resolving power of R$\sim$2000, it provides an ideal dataset
for simulation tests on GSJT data.

The SDSS has provided estimations of the atmospheric parameters for
a subset of their observed stellar spectra since the sixth public
data release (DR6, Adelman-McCarthy et al. 2008). The program to
determine atmospheric parameters ($T_{\rm{eff}}$, log~$g$ and [Fe/H])
for SDSS/SEGUE spectra is called SSPP (Lee et al. 2008a, 2008b),
which is integrated by multiple techniques, based on
medium-resolution spectroscopy and $ugriz$ photometry obtained 
during the course of SDSS-I and SDSS-II/SEGUE. Note that SSPP is
only valid over the temperature range 4500\,-\,7500\,K, and Lee 
et al. (2008a) stated the typical (external) uncertainties as: 
$\sigma$($T_{\rm{eff}}$)\, 
=\,157\,K, $\sigma$(log~$g$)\,=\,0.29\,dex, and $\sigma$([Fe/H])\,
=\,0.24\,dex. Since these reported errors are computed with SDSS
spectra with SNR\,$>$\,50, the real uncertainties of their estimated 
parameters could become larger with declining SNRs 
(as shown in Table 6 of Lee et al. 2008a). Allende Prieto et al. 
(2008) reported high resolution spectroscopy of 126 field stars 
previously observed as part of the SDSS/SEGUE. Using these 
calibration stars, they compared the SSPP determined parameters 
with those derived directly from high resolution spectra of the 
same stars.

To assess the reliability and precision of our determinations
of the atmospheric parameters, we will fit the SDSS spectra
of the high resolution (HR) templates as well as of randomly
selected samples of SEGUE stars in three different SNR ranges.
We will compare the results with the HR  \citep{segueiii}
and SSPP measurements.

\vspace {0.5cm}
{\it 1) Validation from the SDSS High-Resolution Spectroscopy: High Resolution Test}
\vspace {0.5cm}

Out of the 126 SDSS stars with stellar atmospheric parameters
measured from high resolution spectra listed in Allende Prieto
et al. (2008), we selected 125, excluding SDSS J205025.83-011103.8
for which SSPP failed to fit. The high resolution spectra were
obtained by Keck I/HIRES \citep{vogt94}, Keck II/ESI
\citep{shei02}, VLT/UVES, HET/HRS \citep{ram98, tull98}, and
SUBARU/HDS \citep{nog02}; details about the methods for determining
the parameters of these different spectra are described in Allende
Prieto et al. (2008). They empirically determined the typical
random uncertainties of the $T_{\rm{eff}}$, log~$g$, and [Fe/H]
delivered by the SSPP to be 130\,K (2.2\,\%), 0.21\,dex, and
0.11\,dex respectively. They then compared SSPP estimations and those
measurements derived from the high resolution spectra of SDSS
stars. When comparing with the HET data sample which contains 81
stars, the standard deviation of the differences are 2.75\%, 0.25\,
dex and 0.12\,dex for $T_{\rm{eff}}$, log~$g$ and [Fe/H]
respectively, while comparing with data of the other sources
(e.g., Keck, SUBARU), the 1$\sigma$ errors are 3.14\%, 0.46\,dex
and 0.41\,dex respectively, as listed in Table 6 of Allende
Prieto et al. (2008).

\begin{figure}
\includegraphics{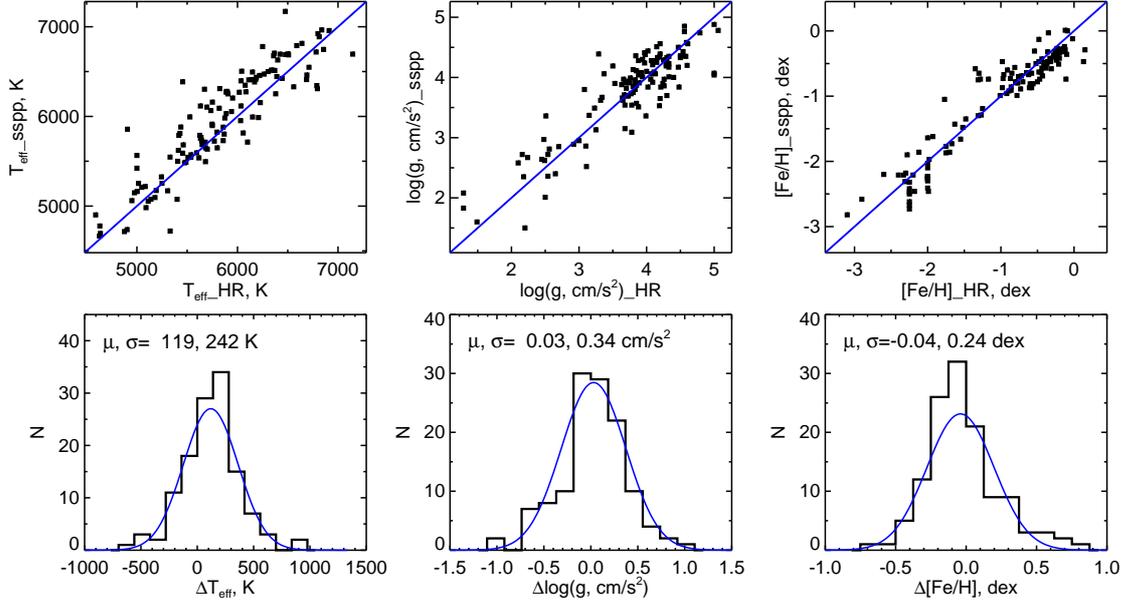} 
\caption {Comparison of the SSPP determined atmospheric
parameters with those derived from the SDSS high resolution
spectra for 125 SDSS observations. The convention is as in
Fig.~\ref{fig.cftest}.
}
 \label{fig.sspp_vs_hr}
\end{figure}

\begin{figure}
\includegraphics{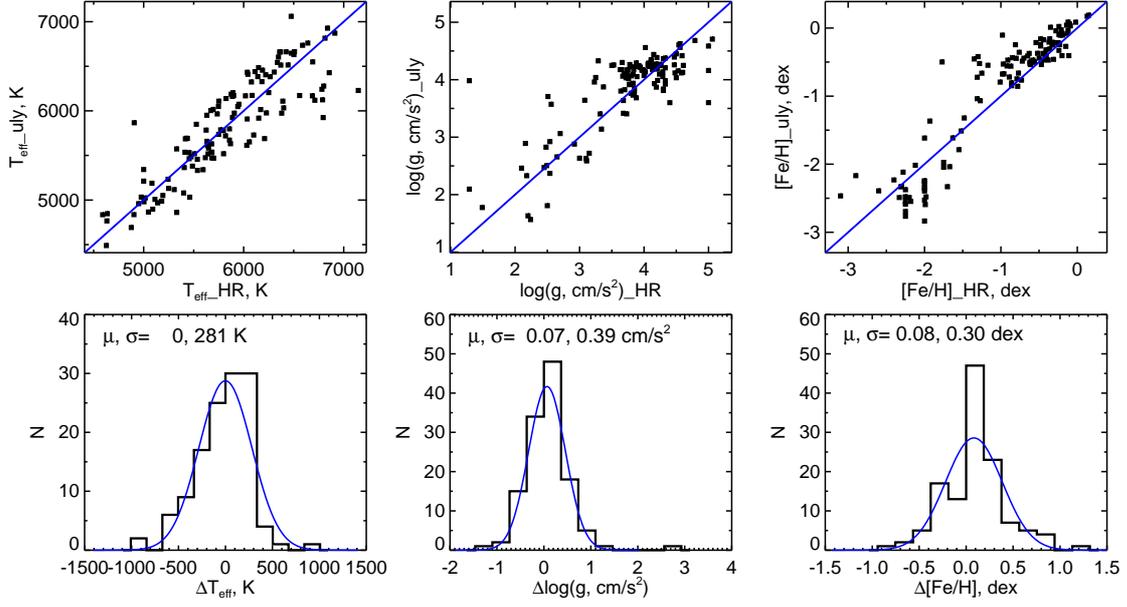}  
\caption {Comparison of the ULySS determined atmospheric
parameters (fitting the SDSS/SEGUE survey spectra) with those 
derived from the SDSS high resolution spectra for 125 SDSS
observations. The convention is as in Fig.~\ref{fig.cftest}.
}
 \label{fig.uly_vs_hr}
\end{figure}

\begin{figure}
\includegraphics{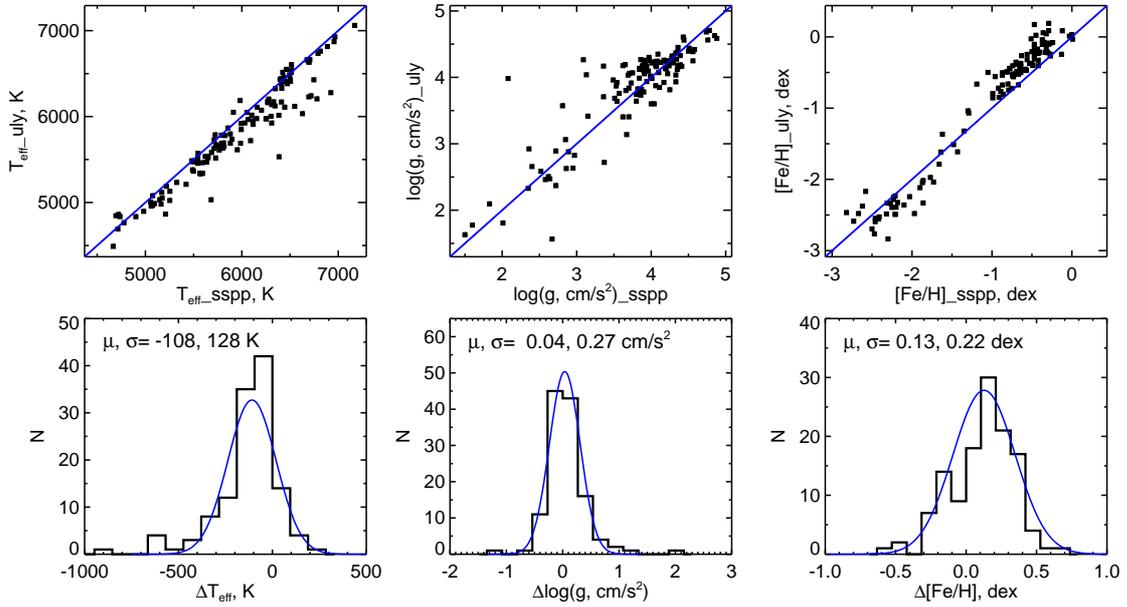} 
\caption {Comparison of the ULySS determined atmospheric
parameters with those given by SSPP for 125 SDSS
observations. The convention is as in Fig.~\ref{fig.cftest}.
}
 \label{fig.uly_vs_sspp}
\end{figure}

Here we made the following comparisons:
\begin{enumerate}
  \item Parameters given by SSPP with the ones derived from the SDSS
HR data;
  \item ULySS determined parameters (fitting the SDSS/SEGUE survey spectra)
with those derived from the SDSS HR data;
  \item ULySS determined parameters with the ones given by SSPP.
\end{enumerate}
Results are sequentially displayed in Fig.~\ref{fig.sspp_vs_hr},
Fig.~\ref{fig.uly_vs_hr}, Fig.~\ref{fig.uly_vs_sspp}, and
Table~\ref{tb.HR}. Unlike Allende Prieto et al. (2008), our 
comparison between SSPP and HR measurements, as 
listed in the top row of Table~\ref{tb.HR}, are averaged values.
It is found that the SSPP estimated temperatures are
systematically 120\,K higher than the HR results; this feature
is comparable with the results specified in Allende Prieto et
al. (2008), where they stated that for the HET sample (number\,
=\,81), SSPP indicated higher $T_{\rm{eff}}$ by about 170\,K
($\sim$3.11\%), and for the other data sample (Keck \& SUBARU,
number\,=\,44), the bias in $T_{\rm{eff}}$ is $\sim$-0.58\% (in
their Sect. 6.2 and Table 6). Fig.5 shows that there is no notable 
offset between the ULySS determined $T_{\rm{eff}}$ and that 
measured from HR spectra. Thus not surprisingly, there is
a 109\,K systematic difference between the ULySS and SSPP
temperatures as shown in Table 1 and Fig.6. There is no obvious
systematic differences in log~$g$ among the three measurements.
As for the metallicity, SSPP measurements are a little bit lower
than the HR results, while ULySS estimations are somewhat higher
than the HR values, with an offset of 0.13\,
dex between ULySS and the SSPP values. For
the 1$\sigma$ errors of the three atmospheric parameters, both
ULySS and SSPP estimations are in an acceptable range, as shown by
comparing with parameters derived from the HR data, and the
ULySS and SSPP values are quite consistent with each other.
This is not unexpected, since both of these methods are processing
medium-resolution spectra in an automated way, which would
result in measurements with more or less similar uncertainties.

\begin{table}
\caption{\label{tb.HR}Comparisons between the Atmospheric
Parameters Derived from the SDSS High Resolution Spectra with those
Determined by ULySS and SSPP for 125 objects.}
\begin{tabular}{c|cc|cc|cc}
\hline\hline
Comparison&\multicolumn{2}{c|}{$\Delta$$T_{eff}$ (K)}&\multicolumn{2}{c|}{$\Delta$log~$g$ ($\rm{cm~s}^{-2}$)}&
\multicolumn{2}{c}{$\Delta$[Fe/H] (dex)}\\
          & $\mu$ & $\sigma$ & $\mu$ & $\sigma$ & $\mu$ & $\sigma$ \\
\hline
SSPP vs. HR    &  120 & 243 &  0.03 & 0.34 &-0.04 & 0.24 \\
ULySS vs. HR   &    0 & 282 &  0.07 & 0.39 & 0.08 & 0.30 \\
ULySS vs. SSPP & -109 & 129 &  0.04 & 0.27 & 0.13 & 0.22 \\
\hline
\end{tabular}
\tablecomments{0.54\textwidth}{The $\mu$ \& $\sigma$ are computed
by the former one minus the latter one, e.g., for the first line,
they are the values of ULySS\,-\,HR parameters.
}
\end{table}

\vspace {0.5cm}
{\it 2) Validation with SEGUE SSPP Determinations: the SNR Test}
\vspace {0.5cm}

For the purpose of validation and calibration of the SSPP output,
only bright stars in the SDSS/SEGUE sample were accessible for
follow-up high resolution spectroscopic observations, thus the
available SDSS high resolution spectra have quite high SNR,
e.g. with SNR\,=\,20\,$\sim$\,80 as indicated in Table 1 of
Allende Prieto et al. (2008). However, most
of the SDSS stellar spectra have significantly lower SNRs,
typically 10\,$\sim$\,20 wavelength-averaged (Sect.7 of Allende
Prieto et al. 2008). By using the SEGUE sample data, we 
have checked the impact of different SNRs on the ULySS 
determined atmospheric parameters.

\begin{figure}
\includegraphics{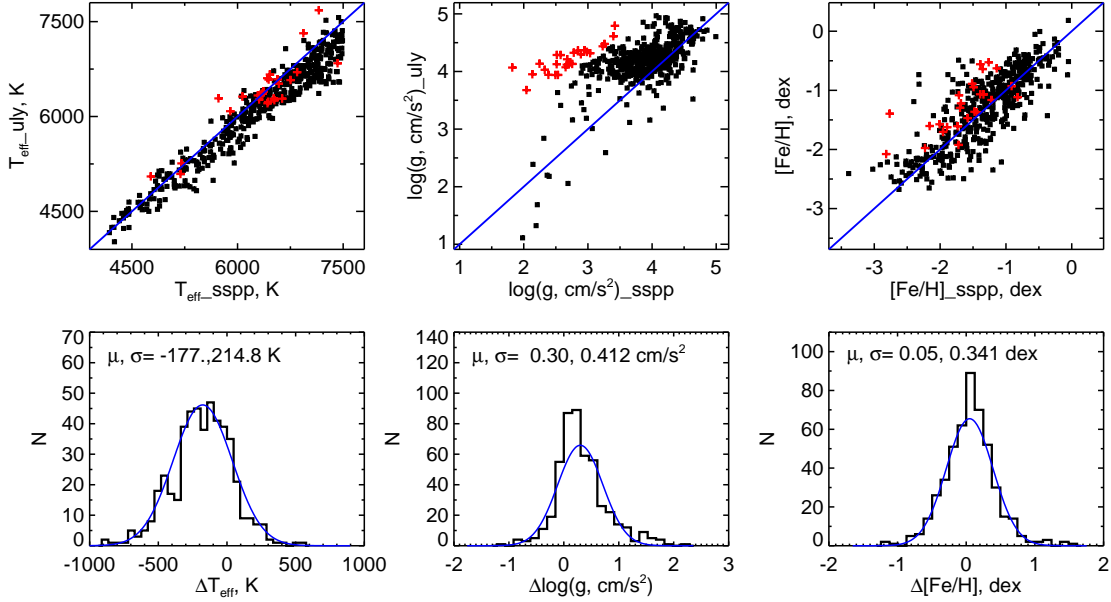}  
\caption {Comparison of the ULySS determined atmospheric
parameters with those computed by the SEGUE SSPP for 486
randomly selected SDSS observations with SNR between 10
and 20. There are 26 (out of 486) log~$g$ discrepant 
measurements, they are displayed in red crosses and 
discussed in the text. 
The convention is as in Fig.~\ref{fig.cftest}.
}
 \label{fig.sn10-20}
\end{figure}

\begin{figure}
\includegraphics{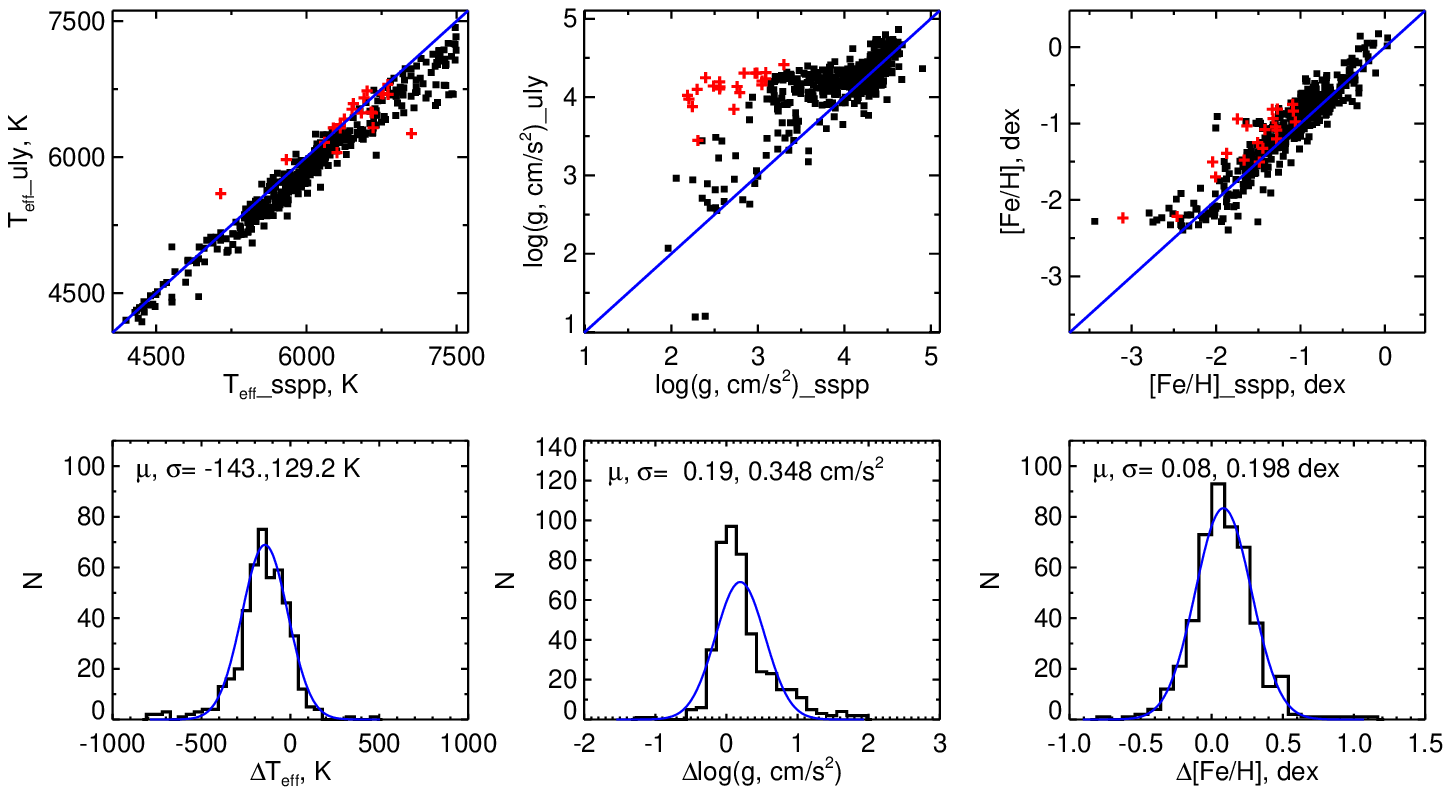}  
\caption {Comparison of the ULySS determined atmospheric
parameters with those computed by the SEGUE SSPP for 466
randomly selected SDSS observations with SRN between 20
and 30. There are 21 (out of 466) log~$g$ discrepant 
measurements, they are displayed in red crosses and 
discussed in the text. The convention is as in 
Fig.~\ref{fig.cftest}.
}
 \label{fig.sn20-30}
\end{figure}

\begin{figure}
\includegraphics{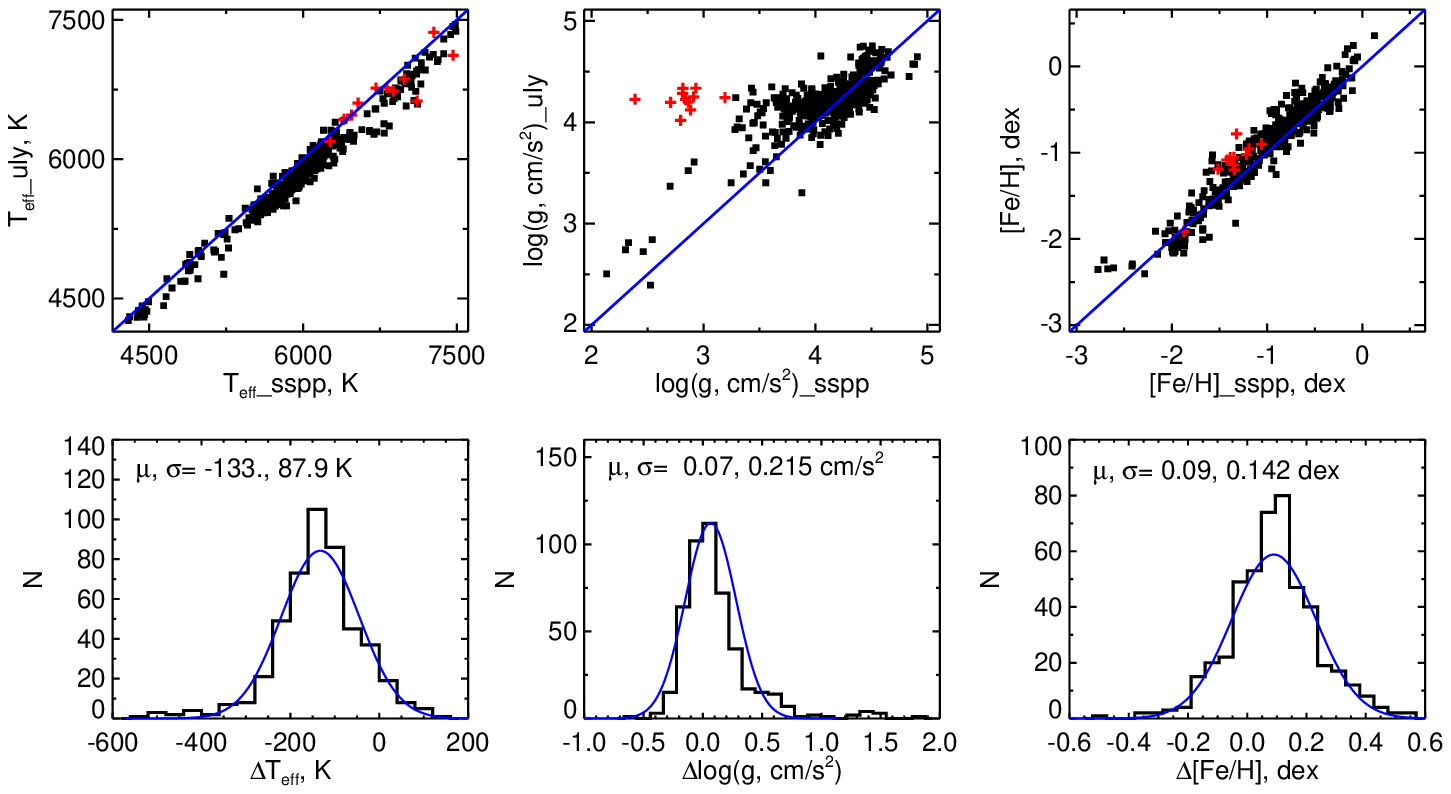}  
\caption {Comparison of the ULySS determined atmospheric
parameters with those computed by the SEGUE SSPP for 476
randomly selected SDSS observations with SNR greater than
30. There are 11 (out of 476) log~$g$ discrepant 
measurements, they are displayed in red crosses and 
discussed in the text. The convention is as in 
Fig.~\ref{fig.cftest}.
}
 \label{fig.sn30}
\end{figure}

We randomly selected some SEGUE spectra together with the
related SSPP estimates for which all the three atmospheric
parameters were available. As mentioned in Lee et al. (2008a), 
the SSPP is reliable in the temperature region cooler than 
7500\,K, so we picked the data with this $T_{\rm{eff}}$ upper 
restriction. We divided them into three groups, with SNRs in 
the region of 10\,$\sim$\,20, 20\,$\sim$\,30 and $>$\,30. Each 
group contains around 475 spectra, mostly F, G, and K type stars. 
The statistics of all the three groups' comparisons are shown 
in Table~\ref{tb.sn}. The comparisons between the ULySS
estimated stellar atmospheric parameters and those from the
SSPP at different SNR regions are displayed sequentially in
Fig.~\ref{fig.sn10-20}, Fig.~\ref{fig.sn20-30}, and
Fig.~\ref{fig.sn30}.

For the three parameters, the mean offsets and the 1$\sigma$ 
errors of the third group are equivalent to the previous 
comparison results between ULySS and SSPP (see Table~\ref
{tb.HR}). Consistently, we can see that the SSPP determined 
$T_{\rm{eff}}$ were underestimated by more than 100\,K when 
compared to the ULySS results. The estimation precision of 
the log~$g$ is to some extent lower than the $T_{\rm{eff}}$ 
and [Fe/H]. 
From Fig.~\ref{fig.sn10-20}, Fig.~\ref{fig.sn20-30}, and 
Fig.~\ref{fig.sn30}, in each figure's log~$g$ panel (upper 
middle one), there exists a strange stripe in the region of 
log~$g$\,$<$\,3.50\,dex, where the SSPP determined log~$g$ 
are systematically smaller than the ULySS 
ones. We marked those discrepant values, with a log~$g$ difference greater 
than 0.90\,dex, as red crosses in 
the three comparison figures. The statistics after excluding 
the discreapant measurements are listed in each group's second line in 
Table~\ref{tb.sn}.

\begin{figure}
\centering
\includegraphics{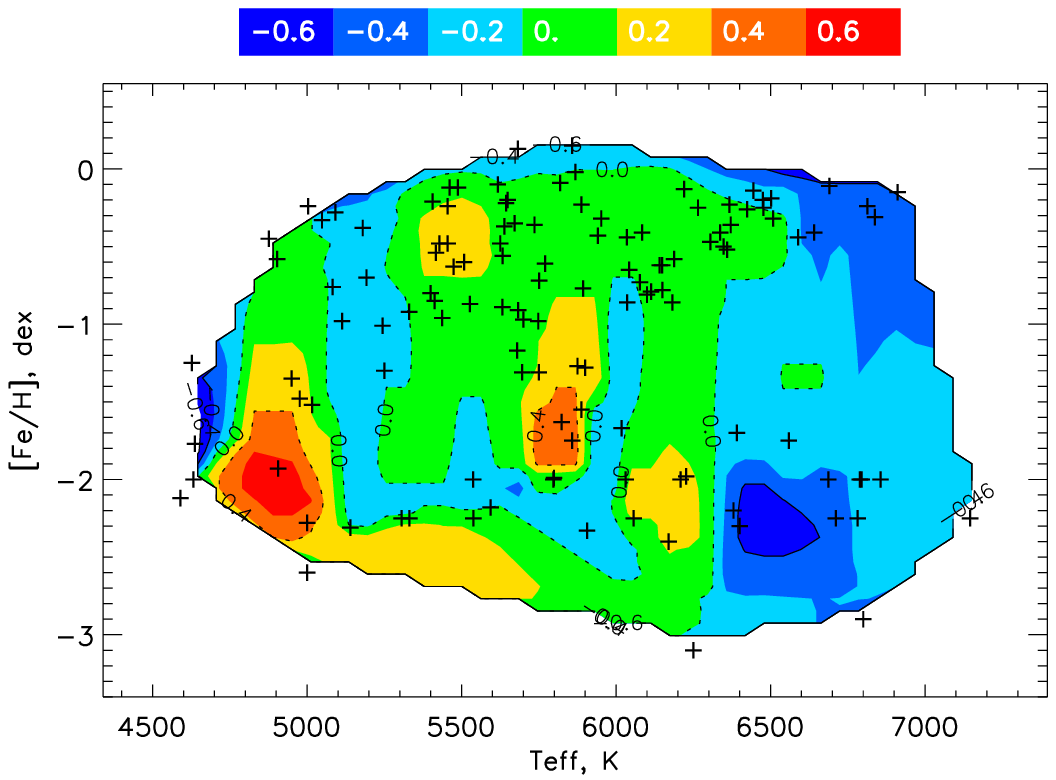}
\caption {Contour map of the differences between the SSPP determined 
log~$g$ and those values derived from the SDSS/SEGUE high 
resolution stellar spectra. The X and Y axes are the effective 
temperature and metallicity values derived from the high resolution 
spectra. The black crosses present the positions where there exist 
actual spectra data.
}
 \label{fig.contour}
\end{figure}

We searched the cause of the discrepancies by 
checking the differences between the SSPP determinations and 
those derived from the SDSS/SEGUE high resolution spectra; the 
result is shown in Fig.~\ref{fig.contour}. This figure is a 
contour map of the differences of the surface gravity 
(SSPP minus HR), while the X and Y axes are the 
$T_{\rm{eff}}$ and [Fe/H] values from the high resolution spectra. 
Apparently, in the region of $T_{\rm{eff}}$ above 6300\,K, the 
SSPP determined surface gravities are systematically 
underestimated by about 0.4\,dex (on average), and this region 
(blue basin) spreads widely in the [Fe/H] space. In the 
solar-like metallicity region, the log~$g$ 
difference could become as big as -0.60\,dex, and the bias area 
can extend to $T_{\rm{eff}}$\,=\,5000\,K. In the metal-poor area 
around [Fe/H]\,=\,-2.30\,dex and $T_{\rm{eff}}$\,=\,6500\,K, the 
log~$g$ difference can be also as big as -0.60\,dex. 
Comparing 
to our detected log~$g$ deviations shown in 
Fig.~\ref{fig.sn10-20}\,$\sim$\,Fig.~\ref{fig.sn30}'s 
temperature panels, those red crosses are likely associated with 
and reside in the blue basin shown in 
Fig.~\ref{fig.contour}; most of them occupy the temperature 
region above 6300\,K, and distributed loosely over the 
metallicity panel. With this investigation, the log~$g$ 
deviations detected in our comparisons between the ULySS and 
SSPP's solutions seem to some extent coming from the internal 
bias of the SSPP pipeline in a specific parameter space --- 
the area of the log~$g$ differences presenting in blue basin shown 
in Fig.~\ref{fig.contour}'s right part. 
This blue basin's mean bias is about 0.40\,dex, 
and we defined the red crosses deviations with log~$g$ 
difference $>$\,0.90\,dex, the magnitude 
of the difference are not exact matches. The 
coutour map is only prepared by using 125 SDSS/SEGUE high 
resolution bright calibration stars (SNR between 20\,$\sim$\,80). 
For the majority of the SDSS/SEGUE fainter survey spectra 
with much lower SNR (10\,$\sim$\,20), the bias area --- the blue 
basin, is 
possibly becoming worse (see statistical examples shown in 
Table 6 of Lee et al. 2008a).
We cannot give any firm conclusion about the origin of the biases, 
but we cannot exclude that they are due to SSPP. This bias
affect 4\% of the sample that we studied, and we 
can consider our determinations as reasonably reliable.

Our tests also show that the precision is
decreasing as the SNR declines, but the determinations do not appear to be biased.

\begin{table}
\caption{\label{tb.sn}Comparisons of the ULySS Determined
Atmospheric Parameters with those computed by SEGUE SSPP
for Some Randomly Selected SDSS Survery Observations with 
Different SNR.}
\begin{tabular}{c|c|cc|cc|cc}
\hline\hline
SNR Group&Num.&\multicolumn{2}{c|}{$\Delta$$T_{eff}$ (K)}&\multicolumn{2}{c|}{$\Delta$log~$g$ ($\rm{cm~s}^{-2}$)}&
\multicolumn{2}{c}{$\Delta$[Fe/H] (dex)}\\
          &      & $\mu$ & $\sigma$ & $\mu$ & $\sigma$ & $\mu$ & $\sigma$ \\
\hline
10\,$\sim$\,20 & 486 & -177 &214.8 &  0.30 &0.412 & 0.05 &0.341 \\
               & 460 & -186 &208.9 &  0.26 &0.357 & 0.03 &0.332 \\ 
20\,$\sim$\,30 & 466 & -143 &129.2 &  0.19 &0.348 & 0.08 &0.198 \\
               & 445 & -147 &125.3 &  0.17 &0.308 & 0.07 &0.191 \\
$>$\,30        & 476 & -133 & 87.9 &  0.07 &0.215 & 0.09 &0.142 \\
               & 465 & -134 & 85.5 &  0.07 &0.211 & 0.09 &0.140 \\
\hline
\end{tabular}
\tablecomments{0.65\textwidth}{The $\mu$ \& $\sigma$ were 
computed by the ULySS estimations minus the SSPP measurements. 
For each comparing group, the first line gives the raw statistics, 
while the second line gives the statistics after excluding the 
log~$g$ outliers which are shown in red crosses in 
Fig.~\ref{fig.sn10-20}, Fig.~\ref{fig.sn20-30}, and 
Fig.~\ref{fig.sn30} respectively.
}
\end{table}

\section{Determining Atmospheric Parameters of GSJT Stars Using ULySS}

\subsection{Determining Strategy for ULySS on the GSJT observations}

We established above that we can get reliable measurements of
the atmospheric parameters from GSJT spectra using ULySS.
Since the model is based on the ELODIE library with a wavelength coverage between
3900~\AA{} and 6800~\AA, we will only use the 
blue arm of the GSJT. Since the method does not require the spectra to be
flux-calibrated, we can use the present commissioning data.

ULySS executes a {\it local} minimization procedure. When approaching 
a new data set with new coverage of wavelength, resolution, parameters, 
etc., one should use the package convergence map tool to study the 
structure of the parameter space and find the region where the local 
minima can potentially trap the solution; see examples in Wu et al. 
(2011a) and Koleva et al. (2009a). For this work, as the targets are 
relatively bright stars, we chose the starting guess grids for the 
three parameters as: $T_{\rm{eff}}$\,=\, [4000, 7000, 15000]\,K, 
log~$g$\,=\,[1.8, 3.8]\,dex, 
and [Fe/H]\,=\,[-1, 0]\,dex. The final solution (absolute minimum) is 
the best from those obtained with different guesses.

Although the blue arm of GSJT covers wavelengths 3700$\sim$5900~\AA,
due to the low instrumental response in the red, we exclude the data 
above 5700~\AA. Moreover, since the ELODIE spectra's SNR drops in the 
blue end, we restricted our fit range to be in 4050$\sim$5700~\AA. 
During the fitting procedure, gaps, bad pixels etc. are automatically 
rejected by the kappa-sigma clips.

To estimate the errors, normally we need to know the random errors of 
each wavelength element. Unfortunately, since the GSJT 2D pipeline is 
still in an early stage of development, it could not provide reliable
errors for this. Thus, following Wu et al. (2011a), we simply
determined an upper limit for these internal errors by assuming that
the residuals purely come from the noise, i.e. the fit is perfect.
With this hypothesis, the reduced $\chi^2$ is by definition equal to
unity. To make this determination, we assumed that the noise was the
same at all wavelength points, and performed the fit with an
arbitrary value of SNR and re-scaled the errors returned by ULySS by
multiplying them by $\sqrt{\chi^2}$. 
To estimate the external errors, we multiplied the internal errors
by a coefficient computed to match the comparisons made with
SDSS/SEGUE (statistics listed in Table 2), and assuming that the 
uncertainties are similar for both series.

\subsection{ULySS Determined Atmospheric Parameters}

\begin{figure*}
\includegraphics{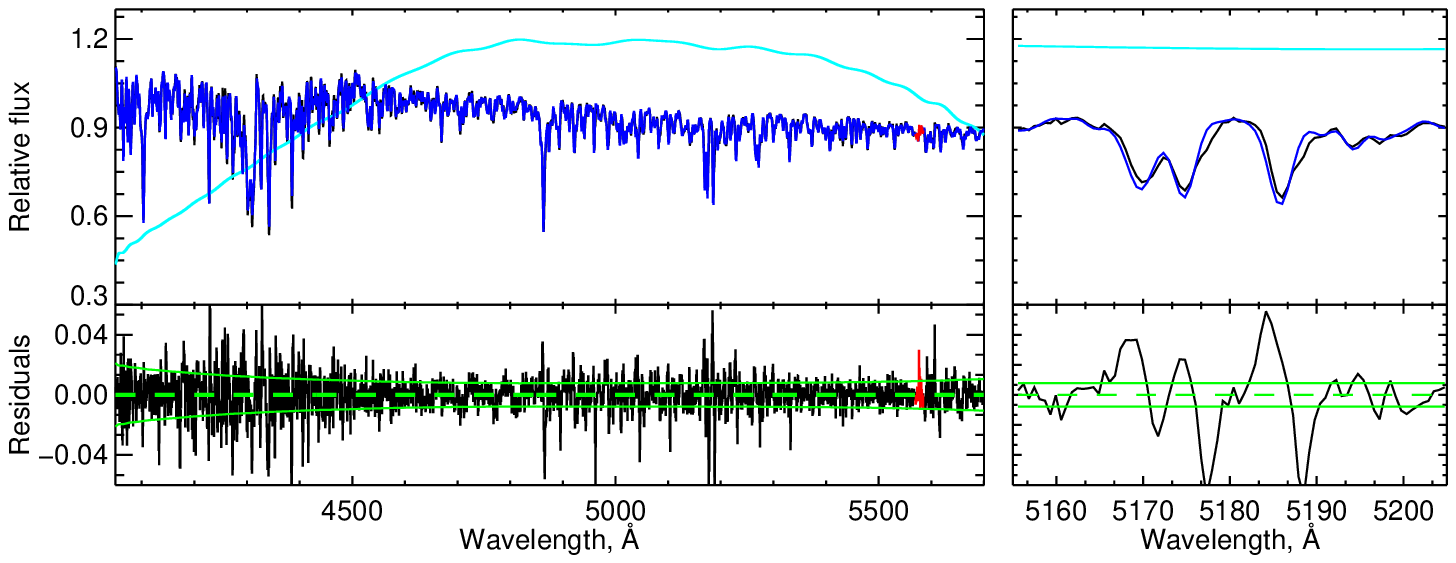}
\caption {ULySS fit of a GSJT observation, obtained with spectrograph
no.\,=\,4, and fiber no.\,=\,193, with a TGM component. The top panel 
shows the spectrum (in black) and the best fit (in blue, both are
almost superimposed and the black line can be seen only when zooming
in on the figure); the light blue is the multiplicative polynomial. In
red, we plot the flagged and masked NaD telluric lines which were not
well calibrated in the ELODIE library. The residuals are plotted in the
bottom panels. The continuous green lines mark the 1-$\sigma$
deviation, and the dashed line is the zero-axis. The right side
expands to a wavelength range around $Mg_{b}$.} 
 \label{fiteg}
\end{figure*}

\begin{figure}
\includegraphics[width=13cm]{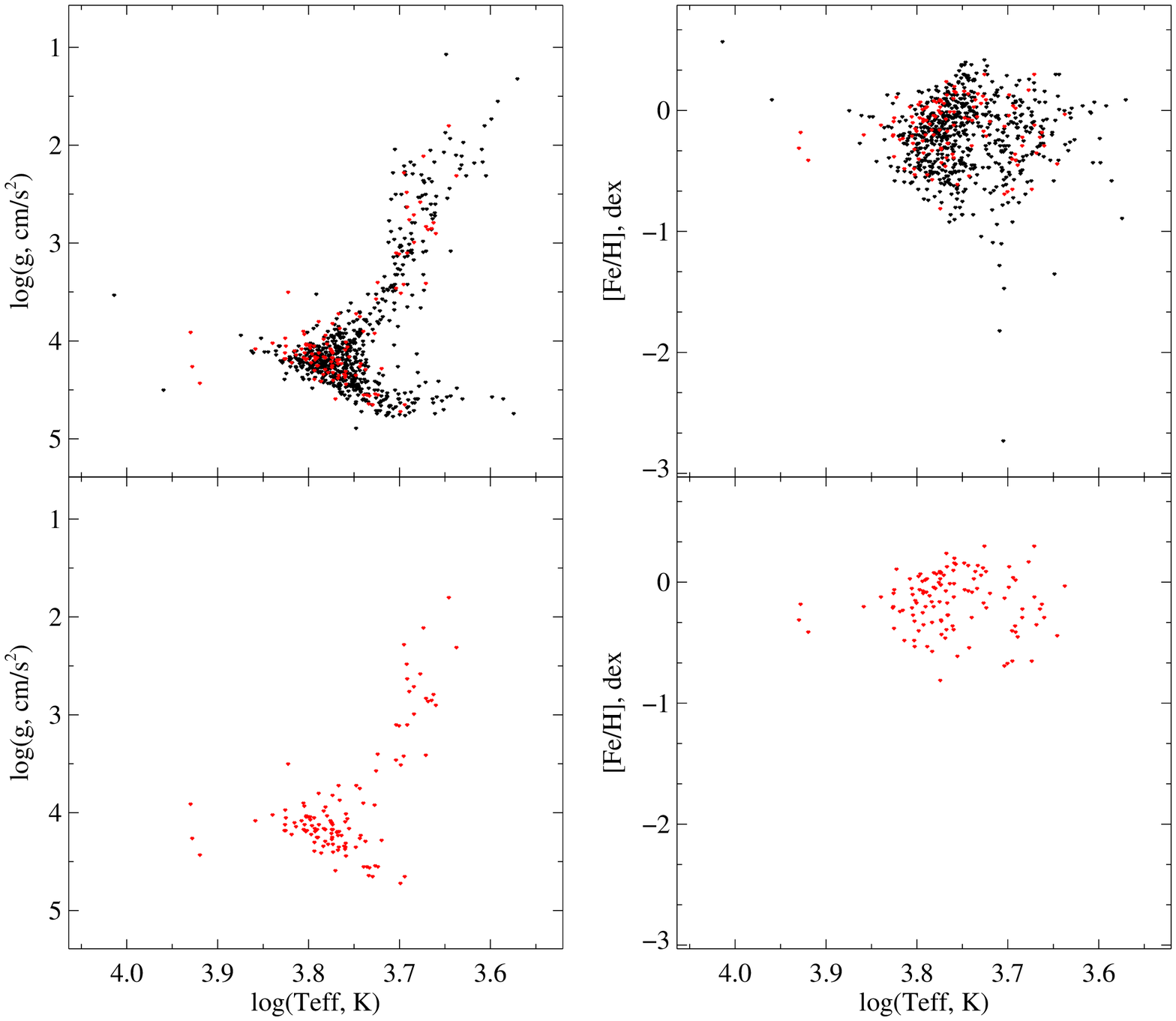}
\caption {Distribution in the log($T_{\rm{eff}}$)\,-\,log~$g$
planes (left) and in the log($T_{\rm{eff}}$)\,-\,[Fe/H] planes (right)
of the adopted atmospheric parameters for the 771 GSJT observed
stars. The 120 red points present the selected empirical templates
for the GSJT. For clarity, they are separately displayed in the
bottom panels.
}
 \label{fig.distri}
\end{figure}

For the selected 771 GSJT stellar spectra from the commissioning
database, we determined their atmospheric parameters by using the ULySS
package. All the estimates are presented in Table~\ref{tb.meas}
(5 objects are shown, the full table is available in the
electronic version), with typical mean internal error bars of
39\,K, 0.21\,dex and 0.11\,dex for $T_{\rm{eff}}$, log~$g$ and 
[Fe/H] respectively. An example of the ULySS fit displayed in
Fig.~\ref{fiteg} illustrates the general quality. The distribution
of all the ULySS determined atmospheric parameters for our GSJT
data sample is shown in Fig.~\ref{fig.distri}. Clearly, all the
solutions are located in the valid estimated parameter space
region as shown in Fig.~\ref{fig.dist_eloI} of the ELODIE library 
(version 3.2). From these results, although there are 
eight metal-poor star candidates (with [Fe/H]\,$<$\,-1.0
\,dex), most of our selected stars assemble around the metal-rich 
region. This phenomenon is restricted by the availability of the 
GSJT commissioning data set, since the commissioning work is 
continuing. In the near future, we will collect and investigate 
more stars with various metallicites.

\subsection{Comparison with Atmospheric Parameters Determined by the SSPP}
 \label{sect:ap}

For an auxiliary check, we compared our determined atmospheric
parameters with those derived by SSPP (version 7.5 via private
communication with T.C. Beers). ULySS determination is fitted 
within the wavelength range of 4050\,-\,5700~\AA, but the SSPP
requires the input spectra to cover the whole wavelength range
in 3800$\sim$9200~\AA. Therefore, we combined the blue and red arm
spectra. Before this step, we visually selected several F8 type
stars in our sample for each spectrograph, and used them to
correct the flux for both the blue and red arm data. Excluding
those targets with missing red arm spectra or with low SNR, 559
out of 771 selected objects were processed with the SSPP.

\begin{figure}
\includegraphics{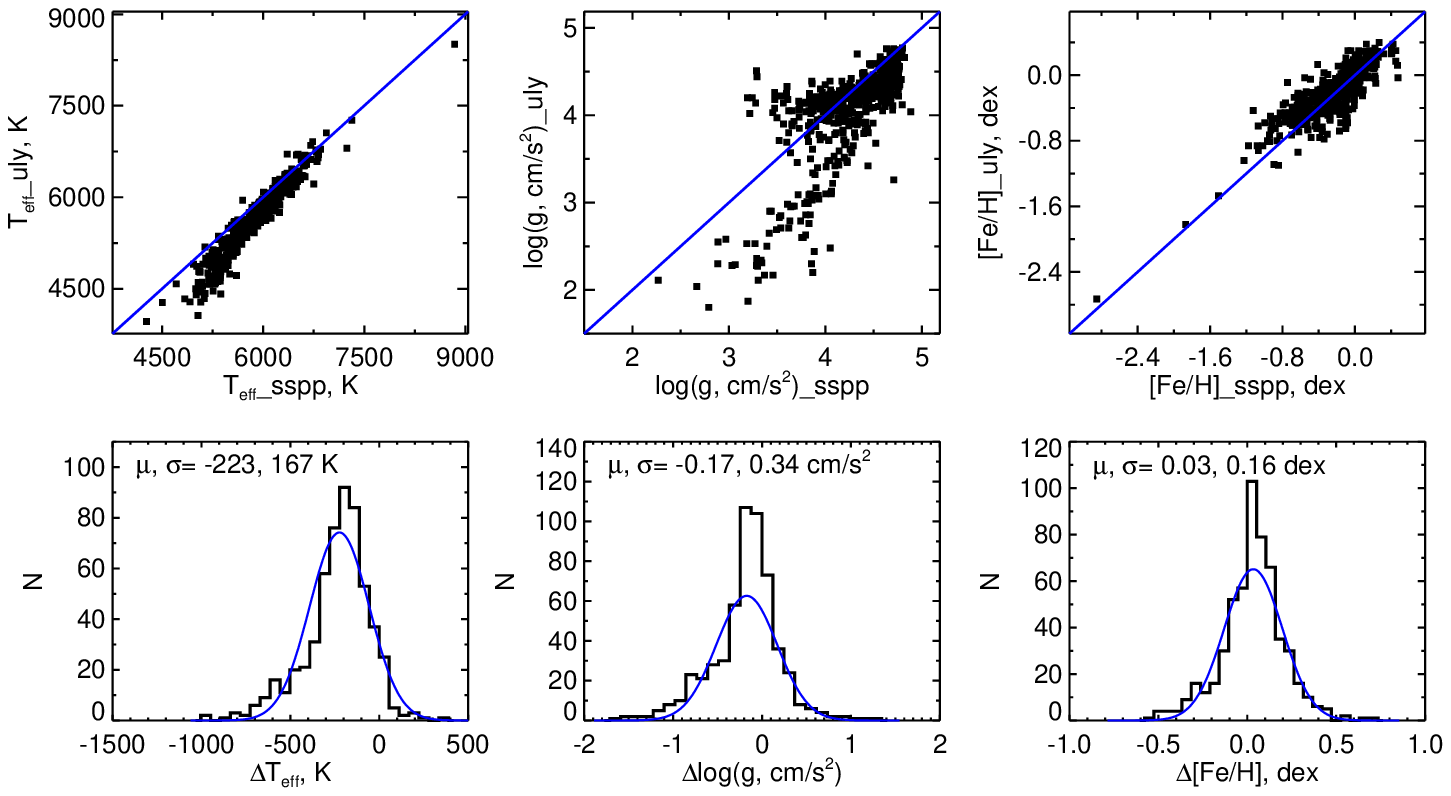}
\caption {Comparison of the ULySS determined atmospheric
parameters with those from SSPP for 559 selected GSJT
observations. The convention is as in
Fig.~\ref{fig.cftest}.
}
 \label{fig.uly_sspp}
\end{figure}

\begin{figure}
\includegraphics[width=15cm]{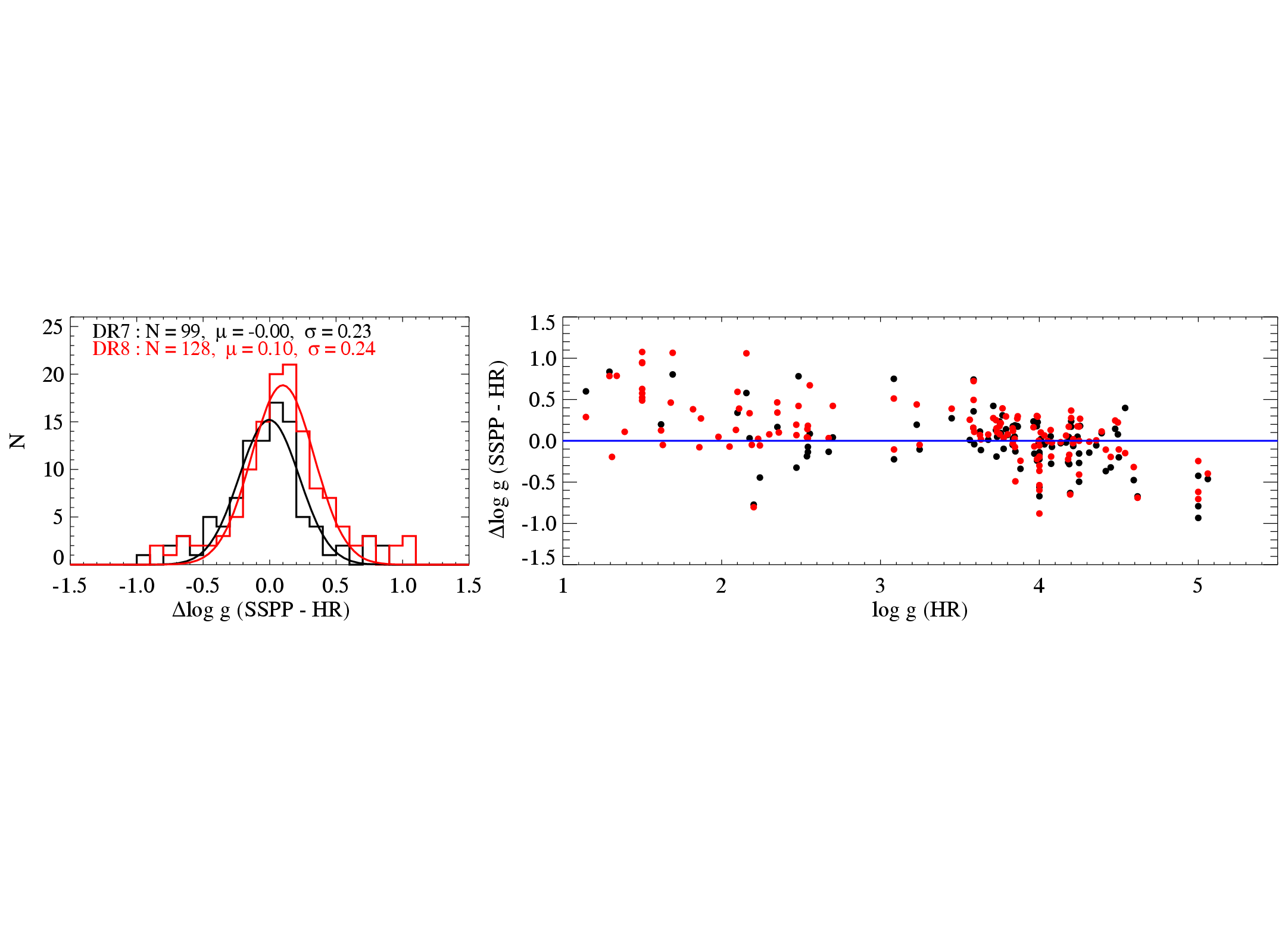}
\caption {Surface gravity comparison between the SSPP version 
7 and the SSPP version 8 for the high-resolution calibration 
stars. The red dots and lines are associated with the SSPP 
version 8, while the black dots and lines correspond to the 
SSPP version 7. This figure is adopted from Fig. A1. of 
Smolinski et al. (2010).
}
 \label{fig.segueiv}
\end{figure}

The measurements of these 559 objects are compared in
Fig.~\ref{fig.uly_sspp} with mean biases (-223\,K, -0.17\,dex,
0.03\,dex) and 1$\sigma$ errors (167\,K, 0.34\,dex, 0.16\,dex)
listed on the bottom panels for each parameter. The -223\,K
systematic bias of the $T_{\rm{eff}}$ is consistent with our
previous validation result for low SNR spectra. 
In the surface gravity panel of the Fig.~\ref{fig.uly_sspp}, 
there is a substructure in the region of log~$g$\,$<3.9$\,dex. 
There the ULySS measurements are systematically underestimated by 
about 0.6\,dex (on average) when compared to the SSPP 
determinations. In Wu et al. (2011a), for the CFLIB library's 958 
F, G and K type stars, ULySS could reach a precision on log~$g$ of 
0.13\,dex and with no significant bias. 
The work of Smolinski et al. 
(2010) could somewhat help us answer this problem. This 
paper is actually one of the series papers coming after Lee et al. 
(2008ab) and Allende Prieto et al. (2008). They enlarged the 
SDSS/SEGUE high resolution stellar spectra number by including 
an extended sample of Galactic globular and open clusters member 
stars. Besides, they announced the SSPP pipeline's recent 
modifications (mainly improved their methods on the 
determination of the [Fe/H] and the radial velocity, for the 
$T_{\rm{eff}}$ and log~$g$ not much) and it has been updated 
from version 7 to version 8. The new version 8 has been applied to 
the SDSS Data Release 8 (DR8) which was scheduled for Dec. 2010. 
By using their high resolution calibration stars, they made surface 
gravity comparison between the SSPP determinations and those values 
derived from the high resolution spectra for both SSPP version 7 
and 8. Their comparison results are displayed in 
Fig.~\ref{fig.segueiv}, version 7 results are shown in black while 
version 8's shown in red. Fig.~\ref{fig.segueiv} clearly shows 
that when log~$g$ is less than 3.9\,dex, on average, the SSPP 
estimations are overestimated by $\sim$\,0.6\,dex when compared to 
the high resolution spectra derivations. This phenomenon explains 
why there is a substructure found in Fig.~\ref{fig.uly_sspp}'s 
log~$g$ panel. The problem is mostly related to the SSPP not 
functioning optimally in this specific log~$g$ region.

The problem 
shown in this section is different from that found in 
Fig.~\ref{fig.sn10-20}, Fig.~\ref{fig.sn20-30}, and 
Fig.~\ref{fig.sn30} (Section of the SNR Test), and these two kind 
of inconsistent cases are not conflicted to each other. 
The log~$g$ problem discussed in the previous context is for stars 
with effective temperature above 6300\,K, and can affect about 
4\% of the data sample. In the log~$g$ panel of 
Fig.~\ref{fig.uly_sspp} we can see several stars appear in 
that bias region. The problem shown in this section is actually 
tending for giants. No apparent visible substructure bias feature 
could be found in Fig.~\ref{fig.sn10-20}\,$\sim$\,Fig.~\ref{fig.sn30}, 
because in that randomly selected data sample, only a few giants 
are included.

For the metallicity, our comparison does not show significant bias.
For the 1$\sigma$ errors of the three parameters, they are in good 
agreement with our validation results listed in Table~\ref{tb.sn}. 
The dispersion on log~$g$ is bigger than the other two parameters, 
though, as mentioned before, log~$g$ is the hardest to estimate 
for the low resolution spectra survey data. Here its larger 
dispersion may partially be due to the low quality of GSJT 
commissioning spectra, i.e., the low SNR, poor flux-calibration 
and simple connection of the blue and red arms, that lead to the SSPP's 
failure in measuring some of our selected spectra. Furthermore, we 
should keep in mind that the SSPP is purposefully tailored and 
adjusted based on SDSS/SEGUE spectra rather than GSJT observations. 
The spectra observed by SDSS/SEGUE and GSJT are different, have 
their own intrinsic characteristic related with each independent 
telescope design, instrumentation and observational conditions etc. 
We would expect the subset offset and large dispersion on surface 
gravity determination will be reduced by both improving the SSPP 
and ULySS capability on the determination of the stellar 
atmospheric parameters, 
as well as the quality of the input GSJT spectral data. Here our 
comparison shows that the two series of the derived parameters 
are consistent with each other on an acceptable scale for 
spectroscopic observations at medium-resolution (R$\sim$2000).

\section{Building the Stellar Spectra Templates for GSJT}

\begin{figure}
\includegraphics{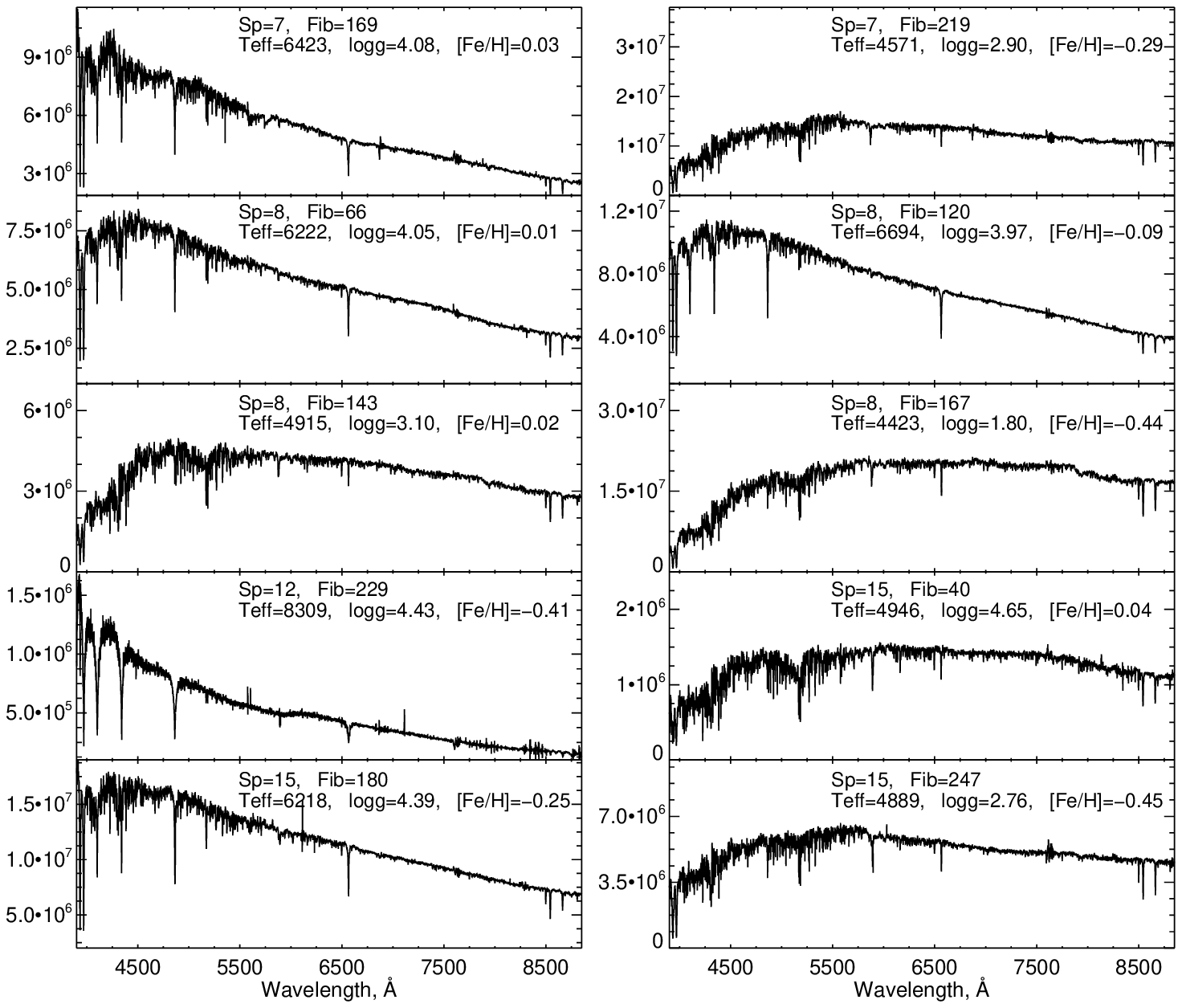}
\caption {Examples of 10 selected empirical stellar spectra
templates of GSJT stars. These spectra have rough 
flux-calibration and non pseudo continuum subtraction. The
relative fluxes are in units of the original counts per pixel.
'Sp' is the spectrograph No.; 'Fib' is the fiber ID No.
}
 \label{fig.temp}
\end{figure}

Our task is aimed at building an empirical set of stellar spectra
templates for GSJT, and at present, we have selected 771 GSJT
observed targets. GSJT practically covers the wavelength range
of 3700\,-\,9000~\AA, but since the flux calibration was not
sufficiently reliable and thus not performed on the spectra, the blue
and red arm spectra were separated. However, for the template
library, it is preferred that the template spectra cover the
full wavelength range. As described in Sect.~\ref{sect:ap}, we
adopted a coarse method to correct the flux and then combined
the blue and red arm spectra. Because of the low quality of the
GSJT commissioning data, and the uncertainties of the flux
calibration and the spectra combination, we selected template
spectra by careful visual inspection, together with
consideration of the atmospheric parameter coverage. Finally,
out of 771 targets, 120 with relatively good quality were
chosen as basic ingredients to build a primary version of the
GSJT stellar spectra template library. These objects are
displayed in red dots in Fig.~\ref{fig.distri} (in both upper
and lower panels) and listed in Table~\ref{tb.meas} (Flag\,=\,1), with
parameters covering the stellar parameter space of
4339$\sim$8507\,K in $T_{\rm{eff}}$, 1.80$\sim$4.72\,dex in 
log~$g$, and -0.81$\sim$0.30\,dex in [Fe/H] respectively.
Most of them are F, G, and K dwarfs. To illustrate them in
more detail, 10 of the sample template spectra are shown in
Fig.~\ref{fig.temp}. We hereby define this newly built GSJT
stellar spectra template library to be Version 1.0.

\section{Summary and Prospects}
 \label{sect:sum}

One of the goals of GSJT surveys 
will be to explore the intrinsic properties of stars in the
Milky Way. To address this, we examine the 
precision of the three fundamental stellar atmospheric parameters 
($T_{\rm{eff}}$, log~$g$ and [Fe/H]) that could be derived from 
the GSJT commissioning observations.

After various validation tests, we demonstrated that ULySS is an
effective tool for measuring large stellar survey spectra at
medium-resolution (R\,$\sim$\,2000). For the first time, we have 
shown that this method can determine the three principal stellar 
atmospheric parameters ($T_{\rm{eff}}$, log~$g$ and [Fe/H]) based 
on GSJT stellar spectra to an accuracy comparable to (or maybe better) 
than that of other measurement approaches like the SSPP. The imbedded
multiplicative polynomial in ULySS, which is used to absorb 
errors in the flux calibration, helps bypass the step of 
normalization to the pseudo-continuum which is usually required 
by other methods. This feature of ULySS is especially favorable 
for GSJT commissioning data, as the flux-calibration has not yet 
been performed.

We selected 771 stars from the GSJT commissioning data
with spectra of relatively good quality. We measured
their parameters with precisions
of 167\,K, 0.34\,dex and 0.16\,dex for $T_{\rm{eff}}$, log~$g$
and [Fe/H] respectively.

This provides us a good opportunity
to construct an empirical stellar spectra template library for
GSJT. At present, this first version consists of 120 stars (the 
best of the 771 selected above). This stellar spectral template 
library for GSJT is valuable and reliable, and it will be used 
as a standard reference frame customized for the GSJT to 
measure the stellar atmospheric parameters using various methods. 
The library will be upgraded in the future, aiming at:

\begin{itemize}
 \item Enlarging the GSJT stellar spectral template library.
   GSJT will continue its commissioning observation, targeting
   a number of open and globular clusters, as well as many
   bright field stars, and it is possible to obtain larger data
   sets for tests and selections. We plan to include more samples
   in the higher and lower temperature region with different
   metallicities. The final goal will be to produce a GSJT
   empirical stellar spectra template library that contains
   several hundred members with wide coverage in the atmospheric
   parameter space.

 \item Assignment of the spectral classifications of the template
   stars. To construct a template library, it would be more
   helpful to include the spectral type of the template star.
   We plan to add approximate MK spectral classification for our
   empirical template stars.
   Methods in
   Singh et al. (1998),
   Bailer-Jones (2002),
   Hawley et al. (2002),
   Covey et al. (2007),
   and Bazarghan \& Gupta (2008)
   may be adopted if they could provide reliable results
   for GSJT observations (Wu et al. 2011b in preparation).

 \item As described in Sect.~\ref{sect:obs&red}, GSJT is still
   in its commissioning period, with problems in the data
   reduction pipeline, e.g., sky-subtraction, flux-calibration, 
   etc., that still need to be solved. With the ongoing 
   adjustments of the instrument and software, we believe that 
   the quality of the GSJT data will be greatly improved, and 
   the templates will be updated accordingly.

\end{itemize}

The newly built GSJT stellar spectra template library (version 1.0)
can be readily deployed in the automated 1D parametrization
pipeline of GSJT. The quality of the template directly relies on an
accurate calibration and pre-processing of the spectra. As the GSJT
commissioning period is still ongoing, future versions of the 
template library will be constructed based on the availability of 
the improved and enlarged GSJT observational database.

Overall, we first effectively determine the stellar atmospheric 
parameters ($T_{\rm{eff}}$, log~$g$ and [Fe/H]) of the GSJT real 
observations (771 homogeneously selected stars) and test the 
estimation precisions on each parameter with external comparisons. 
Moreover, based on this data sample, we select a sub-sample, and 
use these data constructing a preliminary verion of the GSJT 
stellar spectra template library (version 1.0). This work gives 
us an early first light on the scientific capability of the GSJT 
for the research work related with our Milky Way.

\begin{table}
\caption{\label{tb.meas} ULySS Determined Atmospheric Parameters for
the Selected GSJT Stars Observed During its Commissioning Period.}
\begin{tabular}{rr|cc|c|rr|rr|rr|c}
\hline\hline
Sp & Fib & RA & Dec & R Mag & $T_{\rm eff}$ & error & log~$g$ & error & [Fe/H] & error & Flag \\
  &   &   &  &   & \multicolumn{2}{c|}{(K)}&\multicolumn{2}{c|}{($\rm{cm~s}^{-2}$)}&\multicolumn{2}{c|}{(dex)} & \\
\hline
 1&    3 & 132.8668060 &   9.4641514 & 13.74&  4823&$\pm$ 37&  4.63&$\pm$ 0.13& -0.38&$\pm$ 0.13 & 0 \\ 
 1&    6 & 133.0262909 &   9.6342649 & 14.22&  5724&$\pm$ 48&  4.31&$\pm$ 0.19& -0.25&$\pm$ 0.13 & 0 \\ 
 1&   10 & 133.0710449 &   9.6591053 & 13.33&  5295&$\pm$ 41&  4.50&$\pm$ 0.16& -0.32&$\pm$ 0.13 & 0 \\ 
 1&   17 & 133.1125793 &   9.5741863 & 14.79&  5128&$\pm$ 55&  4.66&$\pm$ 0.19& -0.24&$\pm$ 0.18 & 0 \\ 
 1&   20 & 132.8823700 &   9.6176205 & 13.98&  5455&$\pm$ 45&  4.60&$\pm$ 0.16&  0.11&$\pm$ 0.13 & 0 \\ 
...& ... & ...         &   ...       & ...  &  ... & ...    &  ... & ...      &  ... &      ...  & ... \\      
\hline
\end{tabular}
\tablecomments{1.0\textwidth}{
This printed version features only 5 stars atmospheric 
parameters determined by ULySS. The full table which 
includes all the 771 stars is only available in the 
electronic version. 'Sp' is the No. of the spectrography, 
and 'Fib' is the fiber ID number. The third and fourth 
columns are the RA and DEC. The fifth column is the R 
band magnitude. The last column is a flag indicating if 
it is a spectrum selected as a member of the stellar 
spectra template library, '0', not a member of the 
library; '1', a member of the library.
}
\end{table}

\begin{acknowledgements}
The authors are grateful to the anonymous referee for valuable 
comments and Dr. James Wicker for detail amending the language 
of our manuscript. We would like to thank Dr. T.C. Beers for 
sharing us the SSPP (version 7.5), and Dr. Xiaoyan Chen for 
many useful suggestions. 
We thank Fengfei Wang's help on selecting F type stars 
(which were used in the approximate flux calibration).
We acknowledge the grant supports 
given by the Natural Science Foundation of China (NSFC) under 
No.10973021, No.10778626 and No.10933001, the National Basic 
Research Program of China (973 Program) No.2007CB815404, and 
the Young Researcher Grant of National Astronomical 
Observatories, Chinese Academy of Sciences. Yue Wu also 
acknowledges a support from the China Scholarship Council 
(CSC) under grant N$\degr$ 2007104275.

The Guoshoujing Telescope (GSJT) is a National Major Scientific 
Project built by the Chinese Academy of Sciences. Funding for 
the project has been provided by the National Development and 
Reform Commission. The GSJT is operated and managed by the 
National Astronomical Observatories, Chinese Academy of 
Sciences. This work was made with using the SDSS/SEGUE data 
and its CAS database\footnote{\url{http://cas.sdss.org/}}.

Funding for the Sloan Digital Sky Survey (SDSS) and SDSS-II
has been provided by the Alfred P. Sloan Foundation, the 
Participating Institutions, the National Science Foundation, 
the US Department of Energy, the National Aeronautics and 
Space Administration, the Japanese Monbukagakusho, the Max 
Planck Society, and the Higher Education Funding Council 
for England. The SDSS Web site is http://www.sdss.org. The 
SDSS is managed by the Astrophysical Research Consortium 
(ARC) for the Participating Institutions. The Participating 
Institutions are the American Museum of Natural History, 
the Astrophysical Institute Potsdam, the University of 
Basel, the University of Cambridge, Case Western Reserve 
University, the University of Chicago, Drexel University,
Fermilab, the Institute for Advanced Study, the Japan 
Participation Group, The Johns Hopkins University, the 
Joint Institute for Nuclear Astrophysics, the Kavli 
Institute for Particle Astrophysics and Cosmology, the 
Korean Scientist Group, the Chinese Academy of Sciences 
(LAMOST), Los Alamos National Laboratory, the Max Planck 
Institute for Astronomy (MPIA), the Max Planck Institute 
for Astrophysics (MPA), New Mexico State University, Ohio 
State University, the University of Pittsburgh, the 
University of Portsmouth, Princeton University, the 
United States Naval Observatory, and the University of 
Washington.

\end{acknowledgements}

\label{lastpage}

\end{document}